\documentclass[preprint,showpacs,pre,amssymb]{revtex4}

\usepackage{graphicx}
\usepackage{amsmath}
\begin{document}

\title{Stochastic Dynamical Model of Intermittency  in  Fully Developed Turbulence}

\author{Domingos S.~P.~Salazar}
\email{dsps@df.ufpe.br}
\affiliation{Laborat\'orio de F\'\i sica Te\'orica e Computacional,
Departamento de F\'{\i}sica, Universidade Federal de Pernambuco, 50670-901
Recife, PE, Brazil}

\author{Giovani L.~Vasconcelos}
\email{giovani@lftc.ufpe.br}
\affiliation{Laborat\'orio de F\'\i sica Te\'orica e Computacional,
Departamento de F\'{\i}sica, Universidade Federal de Pernambuco,
50670-901 Recife, PE, Brazil}

\begin{abstract}
A novel model of intermittency  is presented in which the dynamics of the rates of energy transfer between successive steps in the energy cascade is described by a hierarchy of stochastic differential equations. The probability distribution of velocity increments  is calculated explicitly and  expressed in terms of generalized hypergeometric functions of the type ${_n}F_0$, which exhibit power-law tails. The model predictions are  found to be in good agreement with  experiments on a low temperature gaseous helium jet.   
It is argued that distributions based on  the functions ${_n}F_0$ might be relevant also for other physical systems with multiscale dynamics.  

\end{abstract}

\maketitle

Our current understanding of fully developed turbulence rests upon two main pillars, namely, the energy cascade, whereby energy is transferred from coarser-scaled structures to the finer, and the phenomenon of intermittency, which results from the fluctuations of the rate of energy transfer. Yet combining these two ingredients into a physically coherent model remains an elusive task.  In his 1941  theory (K41),  Kolmogorov \cite{K41} assumed a constant rate of energy dissipation, which implies Gaussian statistics for the velocity increments, in disagreement with experiments that show heavy-tailed distributions at small scales.
The lognormal model of intermittency proposed by Kolmogorov \cite{K62} in his refined theory based on earlier work by Obukhov \cite{obukhov},
 although  in somewhat good agreement with experimental data, has been criticized under several grounds \cite{novikov,mandelbrot}. 
Several other models of intermittency have been discussed in the literature \cite{mandelbrot,frisch_1978,benzi_jphys_84,MS1987,andrews_1989,yamazaki,she_orszag,benzi_prl_1991,eggers_grossmann,she_leveque}, none of which has been found to be fully satisfactory either with respect to their physical basis or in comparison with experimental data \cite{borgas}.

In this paper we present a new model of intermittency where the fluctuating dynamics of the rates of energy transfer between successive steps in the energy cascade is described by a hierarchy of stochastic differential equations. Under certain reasonable assumptions, an integral expression for the stationary probability density function (PDF), $p(\epsilon_r)$, of the energy flux $\epsilon_r$ at a given scale $r$  is obtained. From the knowledge of $p(\epsilon_r)$, the PDF of the velocity increments  
is then calculated and  expressed in closed form in terms of generalized hypergeometric functions of the type ${_n}F_{0}$, which exhibit power-law tails. 
The model predictions are shown to be in excellent agreement with data from experiments on a turbulent gaseous helium jet for several values of the Reynolds number.  It is also argued that the distributions presented here for the first time
are likely to find applications in other physical systems with multiscale dynamics. 


According to the energy cascade picture of turbulence, energy is injected at the integral scale $L$ and transferred to smaller scales  through a hierarchy of eddies of decreasing size, until it is dissipated by viscous effects at the Kolmogorov length scale $\eta$.  Let  us then denote  by $\epsilon_n$ the rate of energy (per unit mass) transferred to the scale $r=L/b^{n}$
from the scale  $L/b^{n-1}$, 
where $b>1$.
 (Typically, one sets $b=2$ but this is not necessary for our analysis.) 
Because the energy flux $\epsilon_n(t)$ is a fluctuating quantity we seek here to describe its dynamics in terms of stochastic processes. 
On the basis of reasonable physical considerations (see below), we propose that
the dynamics of  $\epsilon_n(t)$  is governed by the following set of stochastic differential equations (SDE):
\begin{equation}
\dot{\epsilon_{i}}=\gamma_{i} (\epsilon_{i-1}-\epsilon_{i}) +k_{i} \epsilon_i \xi_{i}(t),  \quad i=1, ..., n, \label{eq:2}
\end{equation}
where the parameters $\gamma_i$ and $k_{i}$ are assumed to be constant in time and $\xi_i(t)$ are mutually independent white noises. The quantity $\epsilon_{0}$ appearing in  Eq.~(\ref{eq:2}) for $i=1$ represents the rate of energy fed into the system at the integral scale $L$ and is considered fixed.

The terms in the right-hand side of  Eq.~(\ref{eq:2}) have a clear physical interpretation. For instance, the deterministic term  represents the unidirectional coupling between successive steps of the energy cascade. Owing to this coupling, if we were to neglect the fluctuating term in  Eq.~(\ref{eq:2}) then all quantities $\epsilon_i$ would relax   to the constant value $\epsilon_0$, thus recovering the K41 theory. By the same token, Eq.~(\ref{eq:2}) implies that the average energy flux is scale independent, in the sense that in the stationary regime one has $\left<\epsilon_i\right>=\epsilon_0$ for all $i$. The choice of the noise term is also a natural one since we expect a multiplicative noise in a cascade process.
This ensures, in particular, that if the quantities $\epsilon_{i}(t)$ are initially positive then they remain nonnegative for all times.  To see this, note that if $\epsilon_{i}(t)$ were ever to become negative  it would have to cross zero, since it is a continuous process. But if $\epsilon_{i}=0$ at some time, then Eq.~(\ref{eq:2})
implies that  $\dot{\epsilon}_{i}>0$  and so  
$\epsilon_{i}$ will be `reflected' back to the positive range. 
(Of course, the rate of energy transfer $\epsilon_r$ cannot  assume negative values if it is to be identified with the local average rate of energy dissipation, as first suggested by Obukhov \cite{obukhov}.)

The model defined in  Eq.~(\ref{eq:2})  bears some resemblance to shell models of energy cascade in turbulence \cite{biferale_2003}, where one seeks to describe the energy-cascade mechanism by a set of coupled nonlinear ordinary differential equations that are consistent with 
the Navier-Stokes (NS) equation.
Our model is more of a phenomenological nature in that it incorporates the fluctuations of the rates of energy dissipation  explicitly via a set of coupled stochastic differential equations.  
We note, however, that  it is possible  \cite{us}  to give a heuristic derivation  of the deterministic term in  Eq.~(\ref{eq:2}) from the scale-by-scale energy budget equation \cite{frisch}, if one assumes  localness of the energy transfer.
In the same vein, the  noise term can be justified from symmetry considerations
and from the positivity  requirement on   $\epsilon_i$  (see above). A related approach based on energy-balance equations was used in \cite{eggers_1992} to obtain a Langevin description of the energy  of eddies of different sizes, but here the resulting SDE's  are highly nonlinear.  
Our model, in comparison, is written in terms of the energy transfer rate, is linear, and  has the further advantage that it 
 yields an analytical expression for the PDF of the velocity increments which is in excellent agreement with experimental data, as we will see shortly.

Understood in the sense of the It\^o stochastic calculus, Eq.~ (\ref{eq:2}) constitutes a set of $n$ linear SDEs that can be solved exactly  \cite{oksendal}. Such an approach, however, is not very useful for us here since it is not easy  to obtain the stationary joint probability distribution $p(\epsilon_1,...,\epsilon_n)$ from this
exact solution. Considering the stationary Fokker-Planck equation for  $p(\epsilon_1,...,\epsilon_n)$ is not very helpful either, since this equation cannot be easily solved. Thus, an alternative approach is needed to compute the stationary PDF for $\epsilon_n$. 
Here we will take advantage of the separation of the characteristic time scales  at the different steps of the energy cascade \cite{frisch}. To be specific, we make the following assumption:
$\gamma_n^{-1} \ll \gamma_{n-1}^{-1}\ll \cdots \ll \gamma_2^{-1}\ll\gamma_1^{-1}$. 
On the basis of this hypothesis, we can derive the stationary PDF for $\epsilon_{n}$ from our dynamical model, as follows. 


Consider first Eq.~(\ref{eq:2}) for $i=n$. Since the dynamics of $\epsilon_{n}$ has a characteristic time much shorter than that of   $\epsilon_{n-1}$, it is reasonable to assume that before  $\epsilon_{n-1}$ has time to change appreciably  the flux $\epsilon_{n}$ relaxes to a quasi-stationary regime described by a conditional PDF,  $p(\epsilon_n|\epsilon_{n-1})$, obtained assuming $\epsilon_{n-1}$ fixed. In other words, the marginal distribution for  $\epsilon_n$
can be written  as a superposition of  distributions $p(\epsilon_n|\epsilon_{n-1})$ with different values of $\epsilon_{n-1}$: 
$p(\epsilon_{n})=\int_{0}^\infty p(\epsilon_n|\epsilon_{n-1})p(\epsilon_{n-1})d\epsilon_{n-1}$. 
 Implementing this procedure recursively up to the first step of the energy cascade, we obtain
\begin{equation}
 p(\epsilon_n) 
 = \int_{0}^{\infty}\cdots\int_{0}^{\infty} \prod_{i=1}^{n}p(\epsilon_i|\epsilon_{i-1})d\epsilon_1\cdots d\epsilon_{n-1} . \label{eq:pe}
\end{equation}
The distribution $p(\epsilon_i|\epsilon_{i-1})$ can be obtained by solving the stationary Fokker-Planck equation associated with  Eq.~(\ref{eq:2}), holding $\epsilon_{i-1}$ fixed. This yields an inverse-gamma distribution
\begin{equation}
 p(\epsilon_i|\epsilon_{i-1}) = \frac{{(\beta_{i} \epsilon_{i-1})}^{\beta_{i}+1}}{\Gamma (\beta_{i}+1) } {\epsilon_i^{-\beta_{i}-2}}  e^{\frac{-\beta_{i} \epsilon_{i-1}}{\epsilon_i}}, \label{eq:gamma}
\end{equation}
where 
\begin{equation}
\beta_{i}= {2\gamma_i}/{k^{2}_{i}}.
\end{equation}

With the knowledge of the PDF of the energy flux $\epsilon_{n}$, we can  now derive the  PDF  of the longitudinal velocity increments, $\delta _r u=u(x+r)-u(x)$, at a given scale $r$. To this end, we express the marginal distribution for $\delta_r u$
as 
\begin{equation}
P(\delta_r u)=\int p(\epsilon_r) P(\delta_r u|\epsilon_{r})d \epsilon_r,
\label{eq:Pr}
\end{equation}
where  $P(\delta_r u|\epsilon_{r})$ is the conditional probability distribution of $\delta_r u$ for a fixed value of $\epsilon_{r}$. 
Since intermittency  stems from the fluctuations of $\epsilon_r$, it is reasonable to assume that  the statistics of the velocity increments for fixed $\epsilon_r$  is
described by a Gaussian distribution.
This assumption is supported by experiments \cite{naert-1998,SKS}. We then write
\begin{equation}
P(\delta_r u|\epsilon_{r})= \frac{1}{\sqrt{2\pi \sigma^2}}\, \exp\left[-{\frac{(\delta_r{u})^2}{2\sigma^2}}\right]. \label{eq:P1e}
\end{equation}
where $\sigma^2$ is the  (random) variance of $\delta_{r} {u}$ for a fixed value of $\epsilon_{r}$.  In general, one can associate $\sigma^2$ with the local average energy dissipation rate $\epsilon_{r}$ \cite{andrews_1989}. More formally, however, we write 
\begin{equation}
\sigma^2\equiv\langle (\delta_r u)^2| \epsilon_{r}\rangle = \langle (\delta_r u)^2\rangle\, \frac{\epsilon_{r}}{\epsilon_{0}} ,
\label{eq:du2}
\end{equation}
where the second identity  is to be understood in the measure-theoretic sense, meaning that the random variable $\langle \delta u^2| \epsilon_{r}\rangle$ is a coarser version of  $\delta u^2$ \cite{feller}.  

From Eqs.~(\ref{eq:Pr})--(\ref{eq:du2}) it then follows that the PDF  of $\delta_r u$ normalized to unit variance can be written as
\begin{equation}
P(\delta_r \tilde{u}) =\int_{0}^{\infty} \frac{p(\tilde{\epsilon}_{r})}{\sqrt{2\pi \tilde{\epsilon}_r}}\, \exp\left[-\frac{(\delta_r \tilde{u})^2}{2\tilde{\epsilon}_{r}}\right] d\tilde{\epsilon}_r, \label{eq:P1}
\end{equation}
where   $\delta_r \tilde{u}=\delta_r u/\sqrt{\langle (\delta_r u)^2\rangle}$    is the normalized  velocity increment and $\tilde{\epsilon}_{r}=\epsilon_{r}/\epsilon_{0}$ is the normalized energy flux.  
Because we assume a Gaussian of zero mean in  Eq.~(\ref{eq:P1e}), our model describes only the symmetrical part of the PDF of the velocity increments, whose  non-Gaussianity is a signature of intermittency \cite{chevrillard}. The  asymmetry (skewness) of the PDFs is thought to be connected with vortex folding and stretching \cite{chevrillard} and is (for the moment) left out of the model. The idea expressed in Eq.~(\ref{eq:P1}) of writing the PDF of the small-scale velocity fluctuations as a mixture of large-scale (Gaussian) distributions has been used by several authors with different weighting distributions, such as the gamma distribution \cite{andrews_1989}, the lognormal distribution \cite{castaing_PhysD90,chabaud_PRL94,yakhot}, and the chi-square distribution \cite{beck}.
A related description based on a Fokker-Planck equation for 
the conditional distribution of velocity increments was introduced in \cite{peinke_PRL}. In comparision to these previous works, the novelty of our approach is that we model the dynamics of the energy fluxes and then derive (rather than postulate) its distribution, from which the PDF of velocity increments can be obtained explicitly, as shown next.

Upon substituting  Eqs.~(\ref{eq:pe}) and (\ref{eq:gamma})  into Eq.~(\ref{eq:P1}), and performing a sequence of changes of variables,  one can show \cite{us} that
the resulting multidimensional integral can be expressed in terms of known higher transcendental functions:
\begin{equation}
 P(\delta_r \tilde{u}) =
 \frac{1}{\sqrt{2\pi}}\left[\prod_{i=1}^{n}\frac{\Gamma (\beta_{i}+3/2)}{\sqrt{\beta_{i}}\,\Gamma (\beta_{i}+1)} \right] \ _{n}F_{0}(\alpha_{1},...,\alpha_{n};- \frac{(\delta_r \tilde{u})^{2}}{2\beta_{1}\cdots\beta_{n}}),
%
 \label{eq:P5}
\end{equation}
where  $\alpha_i=\beta_i+3/2$ and $_{n}F_{0}(\alpha_{1}, ...,\alpha_{n}; -x)$ is the generalyzed hypergeometric function of order $(n,0)$ \cite{erdelyi}.  
The first two members of the family $_{n}F_{0}$ yield elementary functions, namely,  $_{0}F_{0}$ is the exponential function and $_{1}F_{0}$ is related to the  so-called $q$-exponential: $_{1}F_{0}(1/(1-q),x)=\exp_{q}(x)$,   where $\exp_{q}(x)=[1+(1-q)x]^{1/(1-q)}$. 
We thus see that the distributions  in  Eq.~(\ref{eq:P5}) give a rather natural generalization of the Gaussian ($n=0$) and the  $q$-Gaussian ($n=1$)
distributions. One important property of 
the function $_{n}F_{0}$, for $n>0$, is that it has an asymptotic expansion  \cite{wolfram} of the form
$ _{n}F_{0}(\alpha_{1},...,\alpha_{n};- x) \propto \sum_{i=1}^{n} c_{i}x^{-\alpha_i}\left(1+O(1/x)\right)$, as $x\to\infty$. Thus, the distributions $P(x)$ above comprise a general class of power-law tail distributions with finite variance. (This seems to be the first time that distributions based on  the functions $_{n}F_{0}$, with $n>1$, appear in the literature.)

Returning to our intermittency model given in Eq.~(\ref{eq:2}), we now make the simplifying assumption that  the parameters $\beta_{i}$ are the same throughout the cascade: $\beta_{i}=\beta$. This implies, in particular,  that the distribution $p(\epsilon_i|\epsilon_{i-1})$ given in  Eq.~(\ref{eq:gamma}) is scale invariant in the sense that it has the same functional form regardless of the cascade level. 
In view of the discussion in the preceding paragraph, it then follows that  $P(\delta{u})$ in this case has a single power-law tail: $P(\delta{u})\sim \delta{u}^{-(2\beta+3)}$, for $\delta u\gg 1$.

\begin{figure}[t]
\centering
{\includegraphics[width=0.6\textwidth]{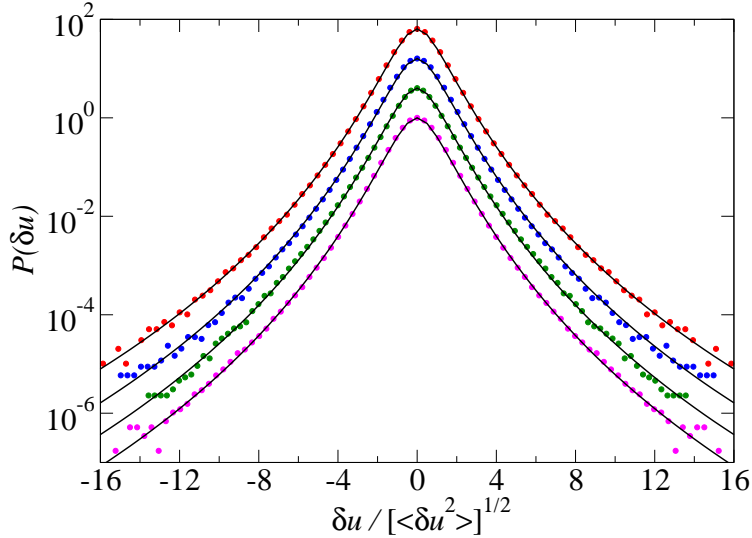}}
\caption{(color online). Histograms of velocity increments (circles) computed from measurements on the axis of a gaseous helium jet for four different Reynolds numbers: $R_\lambda= 463$, 703, 885,  929 (from bottom to top). 
The corresponding solid lines  are the theoretical PDFs for  $n=4$ and $\beta= 6.6$, 6.5, 6.4, 6.2. The curves have been arbitrarily shifted in the vertical direction for clarity.}
\label{fig:1}
\end{figure}

Next we compare the model with experimental velocity measurements on the axis of a low temperature gaseous helium jet.   For details about the experiments the reader is referred to Refs.~\cite{epjb2000,rsi1997}.  
From the recorded data sets,  
typically with $10^{7}$ points each, 
we computed the velocity differences $\delta u$ between two consecutive measurements. 
In Fig.~\ref{fig:1} we show the (symmetrized)  histograms of velocity increments  for four values of the Taylor-scale Reynolds number, namely, $R_\lambda=$ 463, 703, 885, and 929, together with
 the corresponding PDFs  (solid lines) predicted by our model.
The agreement between the theoretical curves and  the experimental data in Fig.~\ref{fig:1} is remarkable. 

\begin{figure}[t]
\centering
{\includegraphics[width=0.6\textwidth]{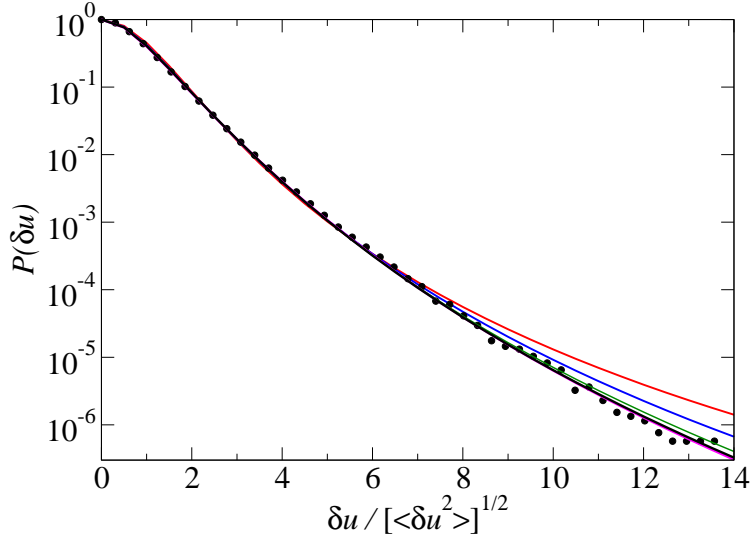}}
\caption{(color online). Distribution of velocity increments for $R_\lambda = 703$  (circles) and corresponding fits (solid lines) for different values of the number $n$ of steps considered in the energy cascade. 
The curves (from top to bottom) are for  $n=1$ and $\beta=1.9$ (red), $n=2$ and $\beta=3.4$ (blue), $n=3$ and $\beta=5.0$ (green), $n=4$ and $\beta=6.6$ (black), and $n=5$ and $\beta=7.9$ (magenta). }
\label{fig:2}
\end{figure}

In Fig.~\ref{fig:1} we chose $n$  as the smallest value necessary to fit satisfactorily the data, in the sense that increasing $n$ gives no further improvement of the fit. This is illustrated in Fig.~\ref{fig:2} where we show the experimental histogram for $R_\lambda=703$, together with the theoretical fits for $n=1, 2, 3, 4, 5$.   Here for each $n$ we chose the value of $\beta$ so as to fit the largest possible range of the data. As $n$ increases, the agreement between the theoretical curve and the data improves considerably, up to a point where  further increasing $n$ makes no practical difference.   Note, in particular, that the $q$-Gaussian ($n=1$)
is in a rather poor agreement with the experimental data, failing most notably to fit the tails \cite{beck-swinney}.
The data shown in Fig.~\ref{fig:2} is for the smallest separation resolved by the experiment, namely, $r=6.5$ $\mu$m, as obtained  from Taylor's frozen turbulence hypothesis. Since  the  Kolmogorov scale  
 in this case is  $\eta =3.4$ $\mu$m  
\cite{epjb2000},
this indicates that our model is apparently valid down to the intermediate dissipation range \cite{frisch}. 
 We have verified  that the model is also able to  fit the PDFs of velocity increments computed at larger separations, with the number  of cascade steps required to fit the data decreasing when $r$ increases, as expected. (More details will be given elsewhere \cite{us}.)

As a final point, let us briefly consider  the structure functions 
predicted by our model.
Using the known properties of the inverse-gamma distribution, one  can show that
\begin{equation}
\langle \epsilon^p_{n}\rangle=\epsilon_{0}^p\left[\prod_{i=1}^{p-1}\frac{\beta}{\beta-i}\right]^n.
\label{eq:enp}
\end{equation}
Recalling that $r=L/b^n$, it then follows  that
the moments of $\epsilon_{r}$ naturally obey a scaling relation
\begin{equation}
\langle \epsilon^p_{r}\rangle\propto\left(\frac{r}{L}\right)^{\tau_{p}},
\label{eq:erp}
\end{equation}
where
${\tau_{p}}= -\sum_{i=1}^{p-1}\log_{b}\left({\beta}/{\beta-i}\right)$.
\label{eq:tau}
Note, however, that the velocity structure functions do not necessarily exhibit scaling, since  $\langle (\delta_r u)^{2p}\rangle=(2p-1)!! \langle (\delta_r u)^{2}\rangle^p \langle\epsilon^{p}_{r}\rangle$ and we impose no  {\it a priori} scaling for the second moment of the velocity increments.
(A more detailed discussion about the scaling properties of our model will be left for a forthcoming publication \cite{us}.)
We also note in passing that our intermittency model 
recovers the lognormal model in the limit of an infinite cascade. 
Indeed, if we take   the limits $n\to\infty$ and $\beta\to\infty$, in such way that $\sigma_r^2\equiv n/\beta$ remains finite,  then  Eq.~(\ref{eq:enp}) becomes
\begin{equation}
\langle \epsilon^p_{r}\rangle=\epsilon_{0}^p e^{\frac{1}{2}\sigma_r^2p(p-1)}, 
\label{eq:enplog}
\end{equation}
which are precisely the moments of a lognormal distribution $\ln {\cal N}(\epsilon_0,\sigma_{r}^2)$ \cite{lognormal}.  Our model thus provides a dynamical context where the lognormal model naturally arises.


In conclusion, we have presented a new cascade model of intermittency in fully developed turbulence based on a hierarchy of stochastic differential equation for  the energy fluxes at different scales in the cascade.  The model  is derived from a physically reasonable set of assumptions and produces an analytical formula for the PDF of the velocity increments in terms of the generalized hypergeometric functions  $_{n}F_{0}$, which fits extremely well  the experimental data. We conjecture that distributions based on $_{n}F_{0}$ are likely to find applications in other systems whose dynamics entails multiple spatial or temporal scales, such as fragmentation processes, biosystems, and financial data.  

\begin{acknowledgments}
 We are grateful to  B. Chabaud and P. E. Roche for providing us with the data. This work was supported in part by the Brazilian agencies CNPq, FINEP,  and FACEPE.
\end{acknowledgments}



\begin{thebibliography}{}
\bibitem{lognormal} A log-normal distribution is not uniquely determined by its moments but this technicality is not relevant here.
\end{thebibliography}


\end{document}